\documentclass[12pt]{article}
\usepackage{amsmath,amssymb,graphicx}

\newcommand{\im}{\text{Im}}
\newcommand{\re}{\text{Re}}
\newcommand{\Tr}{\text{Tr}}
\newcommand{\hc}{\text{h.c.}}
\newcommand{\mev}{\text{MeV}}
\newcommand{\gev}{\text{GeV}}
\newcommand{\tev}{\text{TeV}}
\def\openone{\leavevmode\hbox{\small1\kern-3.8pt\normalsize1}}

\begin{document}

\begin{titlepage}

\begin{flushright}
CERN-TH/2001-016\\
FISIST/02-2001/CFIF\\
\end{flushright}

\begin{center}

\bigskip
{\Large\bf Dynamical \textsl{CP} Violation and Flavour-Changing Processes}\\
\bigskip
\bigskip

{\large G.C. Branco$^{a,b,}$\footnote{E-mail:
gbranco@cfif.ist.utl.pt}, D. Del{\'e}pine$^{c,}$\footnote{E-mail:
David.Delepine@lngs.infn.it} and R. Gonz{\'a}lez
Felipe$^{b,}$\footnote{E-mail: gonzalez@gtae3.ist.utl.pt} }

\bigskip
\emph{$^a$ CERN Theory Division, Geneva, Switzerland\\
$^b$ Centro de F{\'\i}sica das Interac\c{c}{\~o}es Fundamentais,
Departamento de F{\'\i}sica,\\ Instituto Superior T{\'e}cnico,
1049-001
Lisboa, Portugal\\
$^c$ INFN, Laboratori Nazionali del Gran Sasso,\\
I-67010 Assergi (AQ), Italy}

\bigskip

\begin{abstract}

We investigate the phenomenological constraints on a model where,
besides the standard model Higgs sector, there is an effective new
strong interaction acting on the third generation of quarks and
characterized by a $\theta$-like term. This $\theta$ term induces
electroweak symmetry breaking and leads to dynamical spontaneous $CP$
violation. We show that the constraints coming from $K$ physics and
the electric dipole moment of the neutron impose that the new physics
scale should be of the order of 35 TeV. Contrary to naive
expectations, the predictions of the model for $B$ physics are very
close to the standard model ones. The main differences appear in
processes involving the up quarks such as $D^0-\bar{D}^0$ mixing and
in the electric dipole moment of the neutron, which should be close
to the experimental limit. Possible deviations from the standard
model predictions for $CP$ asymmetries in $B$ decays are also
considered.

\vspace{5mm}

\end{abstract}
\end{center}
\end{titlepage}

\section{Introduction}

The fact that the top quark is much heavier than the other quarks,
$m_{t}=174.3 {\pm} 5.1$ GeV \cite{PDG}, is suggestive of a new dynamics
at the electroweak scale, where the third generation may be playing a
special role. In particular, effective four-fermion interactions
\cite{NJL} can lead to the formation of quark-antiquark bound states
which in turn can dynamically trigger the breaking of the electroweak
symmetry \cite{miransky,bardeen}. This is the basic idea of top-quark
condensation as well as of technicolor models, i.e. the Higgs sector
of the standard model is just an effective Ginzburg-Landau-type
description of low-energy physics represented by a composite
isodoublet scalar field (or fields) \cite{cvetic}.

In the above framework, a particularly interesting scenario is
provided by models where the top quark mass arises mostly from a $t
\bar{t}$ condensate, generated by a new strong dynamics, plus a small
fundamental component, generated by an extended technicolor or Higgs
sector \cite{hill}-\cite{gerard}. Such a structure for the top quark
mass avoids the problems usually found in pure (minimal) top-quark
condensation scenarios, which assume that the $t \bar{t}$ condensate
is fully responsible for the electroweak symmetry breaking
\cite{bardeen}, thus leading to a too large $m_t$ value ($m_t \gtrsim
220$ GeV) and a very large scale for the new dynamics ($\Lambda \sim
10^{15}$ GeV) with significant fine tuning.

Along this line, a dynamical scheme was proposed in Ref.
\cite{delepine}, where it is assumed that the third generation of
quarks does indeed experience new forces, symmetric in $t$ and $b$,
and that these new forces also generate a strong $CP$ phase $\theta$.
It is then possible to show that, in such a scenario, the $\theta$
term triggers the breaking of the symmetry between $t$ and $b$ and
induces a large $CP$-violating phase in the Cabibbo-Kobayashi-Maskawa
(CKM) matrix, due to the smallness of the $m_{b}/m_{t}$ mass ratio
\cite{delepine}. In this model one expects to have a richer
low-energy phenomenology when compared to the standard model (SM),
which could lead to potentially interesting effects, specially in $K$
and $B$ physics.

The purpose of this paper is to study the low-energy phenomenological
implications of the model proposed in \cite{delepine} and, in
particular, its implications for $K$ and $B$ physics. We will show
that new observable effects arise due to the fact that the third
generation of quarks experiences new strong forces which in turn lead
to scalar flavour-changing neutral current (FCNC) interactions at
tree level. These FCNC interactions result from the fact that both
the up and down quark mass matrices receive contributions not only
from Yukawa interactions with the standard Higgs but also from
interactions involving the third generation quark-antiquark bound
states.

The present and near future experiments at $B$ factories and the
large hadron collider (LHC) will certainly improve the bounds on many
of the $CP$-violating and flavour-changing processes, which are
forbidden or strongly suppressed in the SM. Therefore it is
particularly interesting to determine possible experimental
signatures in models involving new FCNC physics.

\section{The model} \label{sec:2}

In this section we shall briefly present the main features and
physical consequences of the model in question. A more complete and
detailed analysis can be found in Ref. \cite{delepine}.

We consider a standard model Higgs sector in combination with an
effective new strong interaction acting on the third generation of
quarks and characterized by a $\theta$ term. We require that this new
strong interaction conserves the isospin symmetry between $t$ and $b$
quarks. Moreover, if one assumes that the electroweak symmetry
breaking is induced by radiative corrections due to top-quark (and
possibly, bottom-quark) loops, the quartic self-interactions of the
Higgs field may be neglected. In this case, the relevant classical
Lagrangian for the fundamental scalar field $H$ is given by
\begin{equation}
L_H=D_\mu H^\dagger D^\mu H-m_H^2 H^\dagger H+\left( h_t\bar{\psi}_L
t_R H + h_b\bar{\psi}_L b_R \tilde{H}\ +\hc \right) , \label{eq:1}
\end{equation}
\smallskip where $H=\left(
\begin{array}{l}
H^0 \\
H^-
\end{array}
\right) $ , $\tilde{H}=\left(
\begin{array}{l}
\;H^+ \\
-H^{0^{*}}
\end{array}
\right) $ and $\psi _L=\left(
\begin{array}{l}
t_L \\
b_L
\end{array}
\right)$; $h_t$ and $h_b$ are the Yukawa couplings and $D_\mu$ is the
usual covariant derivative of the SM.

Next one assumes that the interactions acting on the members of the
third generation of quarks are strong enough to form quark-antiquark
bound states at the electroweak scale. The latter can be described in
terms of two complex doublet scalar fields
\begin{equation}
\Sigma_t=\left(
\begin{array}{l}
\Sigma_t^0 \\
\Sigma_t^-
\end{array}
\right) \sim t_R \bar{\psi}_L\;,\qquad \tilde{\Sigma}_b=\left(
\begin{array}{l}
\;\Sigma_b^+ \\
-\Sigma_b^{0^{*}}
\end{array}
\right) \sim b_R \bar{\psi}_L\ ,  \label{eq:2}
\end{equation}
and the corresponding effective Lagrangian then reads:
\begin{equation}
L_\Sigma=D_\mu \Sigma_t^\dagger D^\mu \Sigma_t + D_\mu
\Sigma_b^\dagger D^\mu \Sigma_b-m^2(\Sigma_t^\dagger
\Sigma_t+\Sigma_b^\dagger\Sigma_b)+ g(\bar{\psi}_L t_R\Sigma_t +
\bar{\psi}_L b_R \tilde{\Sigma}_b + \hc)\ .  \label{eq:3}
\end{equation}

The effects of a new strong $CP$ phase $\theta$ can, in principle, be
described through an arbitrary function of $\det U$, where
\begin{equation}
U\sim \left(
\begin{array}{ll}
\bar{t}_L t_R & \bar{t}_L b_{R} \\
\bar{b}_L t_R & \bar{b}_L b_{R}
\end{array}
\right) = \left(
\begin{array}{ll}
\Sigma_t^0 & \;\Sigma_b^{-} \\
\Sigma_t^+ & -\Sigma_b^{0^{*}}
\end{array}
\right) .  \label{eq:4}
\end{equation}

In analogy with QCD $\cite{witten}$ we shall assume the Lagrangian
form\footnote{Another simple choice is given by the 't Hooft
determinant, i.e. $L_\theta = \alpha e^{i \theta} \det U + \hc$.}
\begin{equation}
L_\theta=-\frac{\alpha}{4}\left[ i\Tr \left( \ln U-\ln U^{\dagger}
\right) +2\theta \right] ^{2},  \label{eq:5}
\end{equation}
which typically arises as a leading term in a $1/N$ - expansion.

The total effective Lagrangian of the model is then given by
\begin{equation}
L=L_H+L_\Sigma+L_\theta\ ,  \label{eq:6}
\end{equation}
with $L_H,L_\Sigma$ and $L_\theta$ defined by Eqs.\,(\ref{eq:1}),
(\ref{eq:3}) and (\ref{eq:5}), respectively. Notice that if $h_t=h_b$
the Lagrangian (\ref{eq:6}) conserves an ``isospin'' symmetry.
However, as shown in \cite{delepine}, the angle $\theta$ provides a
dynamical origin for both $CP$ violation and isospin breaking, once
the neutral components of the three doublets $H$, $\Sigma_b$ and
$\Sigma_t$ acquire nonzero vacuum expectation values (VEV's).

Denoting the VEV's of the neutral components of the fields by
\begin{equation}
\left\langle H^0\right\rangle =\frac{v}{\sqrt{2}}\;,\quad
\left\langle \Sigma_t^0\right\rangle
=\frac{\sigma_t}{\sqrt{2}}e^{i\varphi _t},\quad \left\langle
\Sigma_b^0\right\rangle =\frac{\sigma_b}{\sqrt{2}}e^{i\varphi_b}\ ,
\label{eq:7}
\end{equation}
the effective potential reads
\begin{equation}
V=m_H^2\frac{v^2}{2}+\frac{m^2}{2}(\sigma_t^2 + \sigma_b^2)-\beta
\left( \mu_t^2+\mu_b^2\right) + \lambda \left( \mu_t^4+\mu_b^4\right)
+ \alpha \left( \theta - \varphi_t + \varphi_b\right) ^{2}\ ,
\label{eq:8}
\end{equation}
where
\begin{equation}
\mu_i^2 = \frac{1}{2}\left( h_i^2 v^2 + g^2\sigma_i^2 + 2h_i v
g\sigma_i\cos \varphi_i\right)\ ,\ \ \ i=t,b\ ; \label{eq:9}
\end{equation}
$\beta$ and $\lambda$ are some effective quadratic and quartic
couplings, respectively. All couplings and parameters in the
potential are assumed to be real and positive.

The minimization of the potential implies the following system of
equations:
\begin{align}
A_H v &= gh_t I_t\sigma_t\cos \varphi_t +
gh_b I_b\sigma_b \cos \varphi_b\ ,  \nonumber\\
A_t \sigma_t &= gh_t I_t v\cos \varphi_t\ ,  \nonumber\\
A_b \sigma_b &= gh_b I_b v\cos \varphi_b\ ,  \nonumber\\
gh_t I_t v\sigma_t\sin \varphi_t &= -gh_b I_b v\sigma_b\sin \varphi_b
= 2\alpha \left( \theta -\varphi_t + \varphi_b\right)\ ,
\label{eq:10}
\end{align}
where
\begin{align}
A_H &= m_H^2-h_t^2 I_t-h_b^2 I_b\ ,  \nonumber\\
A_i &= m^2-g^2 I_i\ ,  \nonumber\\
I_i &= \beta -2\lambda \mu_i^2\ .  \label{eq:11}
\end{align}
The mass parameters $m_H$ and $m$ are chosen such that the quantities
$A_H$, $A_t$ and $A_b$ defined in Eqs.\,(\ref{eq:11}) are always
positive.

If the parameter $\alpha$ is large, $\alpha \gg \beta m_t^2$, then
the last equation in (\ref{eq:10}) implies the constraint
\begin{equation}
\theta \simeq \varphi_t-\varphi_b\ .  \label{eq:12}
\end{equation}
Furthermore, if $\theta=0$, it is easy to show that
$\varphi_t=\varphi_b=0$ is the only solution of the equations and
therefore $CP$ is conserved.

A simple analytical solution can be given for the isospin symmetric
case $h_t=h_b \neq 0$ and assuming $\beta \gg 2\lambda m_t^2$. In
this case $I_t \simeq I_b\ ,\ A_t \simeq A_b$ and therefore $\sin
2\varphi_t \simeq -\sin 2\varphi_b$. Clearly the large splitting
between the physical values of the bottom and top masses ($m_b \ll
m_t$) requires $\sigma_b \ll \sigma_t$ and thus $\varphi_t \simeq 0,
\varphi_b \simeq -\pi/2$, which in turn demands that the
$CP$-violating phase $\theta$ be close to $\pi/2$. In other words,
the presence of a phase $\theta$ close to $\pi/2$ induces both
isospin breaking and $CP$ violation with

\begin{equation}
\sigma_{b} \ll \sigma_{t} \neq 0\ ,\ v \neq 0\ ,\ \varphi_{t} \simeq
\sigma_{b}/\sigma_{t}\ ,\ \varphi_{b} \simeq -\pi/2 +
\sigma_{b}/\sigma_{t}\ . \label{eq:13}
\end{equation}

The actual values of the VEV's can be determined from the physical
values of the masses $m_b, m_t$ and $m_{W}$. For small values of $v$,
i.e  $v \ll \sigma_{b}, \sigma_t\ $, one has the simple expressions
\begin{equation}
\sigma_b \simeq \frac{m_b v_0}{\sqrt{m_t^2+m_b^2}} \ ,\; \sigma_t
\simeq \frac{m_t v_0}{\sqrt{m_t^2 + m_b^2}}\;, \; \tan \varphi_t
\simeq \frac{m_b}{m_t}\ , \label{eq:14}
\end{equation}
where $v_0 = \sqrt{v^2+\sigma_t^2+\sigma_b^2} = (\sqrt{2}G_F)^{-1/2}
\simeq  246$ GeV, $G_F$ is the Fermi coupling constant.

The mass spectra of the neutral and charged (pseudo) scalars are
easily found. In the neutral sector, it is straightforward to find
the linear combination corresponding to the Goldstone boson eaten up
by the $Z^0$ gauge boson. For $\alpha$ very large, one of the
eigenvalues of the mass matrix will be proportional to
$\sqrt{\alpha}\,$ and therefore the corresponding linear combination
of the fields will decouple from the theory. The remaining $4 {\times} 4$
mass matrix can be easily diagonalized. One finds that the standard
Higgs scalar $h$ has a mass given by $m_h \simeq
2g\sqrt{\lambda}\;m_t\ $, two of the remaining masses are
proportional to $\sqrt{\beta} $ and thus are quite large. Finally the
mass which corresponds mainly to a $\bar{b}\gamma_5 b$ bound state is
very sensitive to the difference $h_t-h_b$, but as soon as $h_t$ and
$h_b$ differ (as expected from higher order corrections) it will also
get a contribution proportional to $\sqrt{\beta}$. In the charged
sector one of the eigenstates is eaten up by the $W$ gauge boson
through the usual Higgs mechanism. For the isospin symmetric case
$h_b=h_t$, we find that one of the charged Higgs masses is very
small, i.e. a new pseudo-Goldstone boson appears as it happens in the
neutral sector. Nevertheless, radiative corrections yield $h_b \neq
h_t$ and therefore this mass will get a large contribution
proportional to $\sqrt{\beta}$.

To conclude this section let us comment on the origin of $CP$
violation in the present model. As shown in Ref. \cite{delepine} the
new interaction characterized by a $\theta \neq 0$ term induces a
$CP$-violating effect which filters down to the SM only if
$m_b/m_t\neq 0$. Moreover, this new source of $CP$ violation can be
in principle responsible for what is observed in the $K^0-\bar{K}^0$
system, since it leads indeed to a sizeable $CP$-violating phase in
the CKM matrix, $\delta_{KM} \simeq -(\varphi_t+\varphi_b) \simeq
\pi/2$.

\section{The structure of flavour-changing interactions}
\label{sec:3}

In general, the presence of more than one Higgs doublet in the SM
leads to FCNC interactions at the tree level, which are mediated by
the physical neutral scalars. Such interactions are severely
constrained by the smallness not only of the $CP$-violating parameter
$\varepsilon_K$ but also of the $K^0-\bar{K}^0$ and $B^0-\bar{B}^0$
mixing. The model we are considering is effectively equivalent to a
three Higgs doublet model with a specific structure for Yukawa
couplings. It is therefore straightforward to determine the form of
the induced FCNC interactions by generalizing the results obtained in
the two Higgs doublet case \cite{branco}.

Let us consider 3 Higgs doublets $\Phi_j$ and make the decomposition:
\begin{equation}
\Phi_j=e^{i\alpha_j} \left(
\begin{array}{c} \phi^+_j \\ \\
\frac{1}{\sqrt{2}}(v_j+R_j+iI_j)
\end{array} \right),\ \ j=1,2,3\ ,
\end{equation}
where $R_j, I_j$ are real fields and $v_j e^{i\alpha_j}$ denote the
VEV's of the Higgs fields. The Yukawa couplings of the Higgs fields
to the quark weak eigenstates are given by
\begin{equation}
L_Y = -(\bar{u}_L \bar{d}_L) \Phi_1 g_1^d d_R -(\bar{u}_L \bar{d}_L)
\Phi_2 g_2^d d_R  -(\bar{u}_L \bar{d}_L) \tilde{\Phi}_1 g_1^u u_R
-(\bar{u}_L \bar{d}_L) \tilde{\Phi}_3 g_3^u u_R\ ,
\end{equation}
where $\tilde{\Phi} \equiv i \sigma_2 \Phi^*$ and $g_i^{u,d}\,
(i=1,2,3)$ are the Yukawa coupling matrices. The quark mass matrices
are easily obtained,
\begin{equation}
M_u = \frac{1}{\sqrt{2}} v_1 g_1^u +\frac{1}{\sqrt{2}} e^{-i\alpha_3}
v_3 g_3^u\ , \label{Mu}
\end{equation}
\begin{equation}
M_d = \frac{1}{\sqrt{2}} v_1 g_1^d +\frac{1}{\sqrt{2}} e^{i\alpha_2}
v_2 g_2^d\ , \label{Md}
\end{equation}
where the phase $\alpha_1$ has been put equal to zero by an
appropriate redefinition of the fields.

To single out the pseudo-Goldstone boson $G^0$ we introduce the new
fields $\phi^0, R$, $R^{'}$, $G^0$, $I$ and $I^{'}$ defined through
the transformation
\begin{equation}
\left(\begin{array}{c}
R_1 \\
R_2 \\
R_3
\end{array}
\right) = O \left(\begin{array}{c}
\phi^0\\
R\\
R^{'}
\end{array}
\right)\ ,\quad \left(\begin{array}{c}
I_1 \\
I_2 \\
I_3
\end{array}
\right) = O \left(\begin{array}{c}
G^0\\
I\\
I^{'}
\end{array}
\right)\ ,
\end{equation}
with
\begin{equation}
O = \left(
  \begin{array}{ccc}
    v_1/v_0 & v_2/v^{'} & v_1v_3/v_0 v^{'} \\
    v_2/v_0 & -v_1/v^{'} & v_2v_3/v_0 v^{'}\\
    v_3/v_0 & 0 &-v^{'}/v_0
  \end{array}
  \right)\label{matrixO}
\end{equation}
and $ v_0^2=v_1^2+v_2^2+v_3^2\ $, $ v^{'2}=v_1^2+v_2^2\ $. In terms
of the new fields, the scalar couplings to the down quarks can be
written as
\begin{align}
L_Y^d &= -\frac{1}{v_0}\bar{d}_L M_d\,d_R (\phi^0+i G^0) -\bar{d}_L
\left(g_1^d \frac{v_2}{v^{'}}-g_2^d
\frac{v_1}{v^{'}}e^{i\alpha_2}\right)d_R \frac{R + i I}{\sqrt{2}}
\nonumber\\ &  -\frac{v_3}{v_0 v^{'}}\bar{d}_L M_d\,d_R (R^{'} + i
I^{'})+\hc\ .
\end{align}

We notice that the couplings to the fields $\phi^0, G^0, R^{'}$ and
$I^{'}$ are flavour-conserving while the couplings to $R$ and $I$ are
flavour-violating. Similarly, for the couplings to the up quarks one
obtains
\begin{align}
L_Y^u &= -\frac{1}{v_0} \bar{u}_L  M_u\,u_R (\phi^0 - i G^0)
-\bar{u}_L \left(g_1^u \frac{v_2}{v^{'}} \right)
u_R \frac{R - i I}{\sqrt{2}}  \nonumber\\
& - \bar{u_L}  \left( g_1^u \frac{v_1 v_3}{v_0 v^{'}}-g_3^u
\frac{v^{'}}{v_0} e^{-i \alpha_3} \right) u_R \frac{R^{'} -i
I^{'}}{\sqrt{2}} +\hc \ ,
\end{align}
and thus the couplings of $\phi^{0}, G^{0}$ conserve flavour while
the couplings of $R, R^{'}, I, I^{'}$ do violate flavour.

It is useful to obtain the scalar-quark couplings in terms of the
quark mass eigenstates. In the down quark sector we find
\begin{equation}
L_Y^d = -\frac{1}{v_0}\bar{d}_L D_d\,d_R (\phi^0+i G^0) -\bar{d}_L
 N_d d_R \frac{R + i I}{\sqrt{2}} -\frac{v_3}{v_0 v^{'}}\bar{d}_L
D_d\,d_R (R^{'} + i I^{'})+\hc\ , \label{LYd}
\end{equation}
where $ D_d = U_{dL}^{\dagger} M_d\,U_{dR} = \text{diag}
(m_d,m_s,m_b)$ and
\begin{align}
N_d &= U_{dL}^{\dagger} \left(g_1^d \frac{v_2}{v^{'}} -g_2^d
\frac{v_1}{v^{'}} e^{i\alpha_2}\right) U_{dR} =
\frac{\sqrt{2}\,v_2}{v^{'} v_1} D_d-\frac{v^{'}}{v_1} e^{i\alpha_2}
G_2^d \ , \label{Nd} \\
G_2^d &\equiv U_{dL}^{\dagger}g_2^d U_{dR}\ . \label{Gd}
\end{align}
In the up quark sector the couplings to the scalars in the quark mass
eigenstate basis are given by
\begin{equation}
L_Y^u= -\frac{1}{v_0} \bar{u}_L D_u\,u_R (\phi^0 - i G^0)-\bar{u}_L
N_u u_R \frac{R - i I}{\sqrt{2}} -\bar{u}_L N_u^{'} u_R \frac{R^{'}
-i I^{'}}{\sqrt{2}} +\hc \ ,
\end{equation}
where $ D_u = U_{uL}^{\dagger} M_u\,U_{uR} =
\text{diag}(m_u,m_c,m_t)$ and
\begin{align}
N_u &= \frac{v_2}{v^{'}}\, U_{uL}^{\dagger} g_1^u\, U_{uR}\ ,\\
N_u^{'} &= U_{uL}^{\dagger} \left(\frac{v_1 v_3}{v_0 v^{'}} g_1^u  -
\frac{v^{'}}{v_0} g_3^u e^{-i\alpha_3}\right)\, U_{uR}\ ,
\end{align}
which can be rewritten as
\begin{align}
N_u &= \frac{v_2}{v_1^2} \sqrt{2} D_u -
\frac{v_2 v_3}{v_1^2} e^{-i\alpha_3} G_3^u \ ,\label{Nu}\\
N_u^{'} &= \frac{v_3}{v_0 v^{'}}\sqrt{2} D_u - \left(
\frac{v_3^2}{v_0 v^{'}} + \frac{v^{'}}{v_0}
\right) e^{-i \alpha_3} G_3^u\ ,\\
G_3^u &\equiv U_{uL}^{\dagger}\, g_3^u\,  U_{uR}\ . \label{Gu}
\end{align}

The Yukawa coupling matrices $g_2^d, g_3^u$ have a very simple form
in the present model, namely,
\begin{equation}
g_2^d = \left(\begin{array}{ccc}
0 & 0 & 0 \\
0 & 0 & 0 \\
0 & 0 & g_b
\end{array}
\right) \ , \quad g_3^u = \left(\begin{array}{ccc}
0 & 0 & 0 \\
0 & 0 & 0 \\
0 & 0 & g_t
\end{array}
\right)\ , \label{gmatrices}
\end{equation}
and therefore the matrices $G_2^d, G_3^u$ defined in Eqs.\,(\ref{Gd})
and (\ref{Gu}) are given by

\begin{align}
  \left(G_2^d\right)_{ij} &= g_b \left(U_{dL}^\ast\right)_{3i}
  \left(U_{dR}\right)_{3j}\ ,\label{G2d}\\
  \left(G_3^u\right)_{ij} &= g_t \left(U_{uL}^\ast\right)_{3i}
  \left(U_{uR}\right)_{3j}\ . \label{G3u}
\end{align}
These matrices completely determine the structure of tree-level FCNC
interactions in the model. Without further assumptions we cannot
predict the size of such interactions. We shall assume that the quark
mass matrices $M_{u,d}$ are hermitian\footnote{According to the polar
decomposition theorem, the mass matrices $M_{u,d}$ can always be
written as a product of a hermitian matrix and a unitary matrix. The
latter can be rotated away by a redefinition of the right quark
fields. Notice however that the form of the coupling matrices $g_2^d$
and $g_3^u$ given in Eq.\,(\ref{gmatrices}) is in general not
invariant under such a transformation. Here we shall assume that the
quark mass matrices are hermitian in the basis where the couplings
have the special form (\ref{gmatrices}). Our analysis can be easily
extended to a more general case.} and that the CKM mixing matrix $V
\equiv U_{uL}^{\dagger} U_{dL}$ is dominated by $U_{dL}$, i.e.
$U_{uL} \simeq \openone$, as favoured phenomenologically. Under the
above ``reasonable'' assumptions, the off-diagonal elements of $N_d$
in Eq.\,(\ref{Nd}) are entirely predicted in terms of $V$ since from
Eq.\,(\ref{G2d}) we obtain

\begin{equation}
  \left(G_2^d\right)_{ij} = g_b V^\ast_{3i} V_{3j}\ . \label{G2da}
\end{equation}

Finally, new contributions to flavour-changing processes will be also
induced by the couplings of the heavy charged Higgs fields to the
quarks. Such contributions correspond to Feynman box diagrams with
$W$-boson and charged Higgs particle exchanges. To determine the
magnitude of these couplings, let us introduce the new charged fields
$G^+, H_1^+, H_2^+$ through the decomposition
\begin{equation}
\left(\begin{array}{c}
\phi_1^{+}\\
\phi_2^{+}\\
\phi_3^{+}
\end{array}
\right)= O \left( \begin{array}{c}
G^+\\
H_1^+ \\
H_2^+
\end{array}
\right)\ ,
\end{equation}
where the matrix $O$ is given by Eq.\,(\ref{matrixO}) and $G^+$
corresponds to the pseudo-Goldstone boson. Going to the physical
basis for the charged Higgs fields, the couplings to the $d_R$ and
$u_R$ quarks are given by
\begin{align}
L_{Y}^{+} = &- \sqrt{2}\, \bar{u}_L M_d\, d_R\, G^{+} - \bar{u}_L
A_1^d d_R H_1^{+} - \bar{u}_L A_2^d d_R H_2^{+} \nonumber\\
&- \sqrt{2}\, \bar{d}_L M_u\, u_R\, G^{-} - \bar{d}_L A_1^u u_R
H_1^{-} - \bar{d}_L A_2^u u_R H_2^{-} + \hc \ ,
\end{align}
where we have introduced the coupling matrices
\begin{align}
& A_i^u = e^{-i \alpha_2} \left(g_1^u O_{1(i+1)}+ g_3^u e^{-i
\alpha_3}O_{3(i+1)}\right)\ ,\\
& A_i^d = e^{i \alpha_3} \left(g_1^d O_{1(i+1)}+ g_2^d e^{i
\alpha_2}O_{2(i+1)}\right)\ ,\quad i=1,2\ .
\end{align}

After performing a rotation to the quark mass eigenbasis we obtain
\begin{align}
&A_i^u= V^{\dagger}e^{-i \alpha_2} \left[ \frac{\sqrt{2}}{v_1} D_u
O_{1(i+1)}+ G_3^u e^{-i\alpha_3} \left( O_{3(i+1)}-\frac{v_3}{v_1}
O_{1(i+1)}\right) \right]\
,\label{Au}\\
&A_i^d=  V\, e^{i \alpha_3} \left[ \frac{\sqrt{2}}{v_1} D_d
O_{1(i+1)}+ G_2^d e^{i\alpha_2} \left( O_{2(i+1)}-\frac{v_2}{v_1}
O_{1(i+1)}\right) \right]\ ,\label{Ad}
\end{align}
with $G_2^d\, ,\, G_3^u$ defined in Eqs.\,(\ref{Gd}) and (\ref{Gu}),
respectively.

It is clear that in order to analyze the charged Higgs contributions
to the relevant flavour-changing processes we need to know the
structure of the unitary matrices $U_{uL},U_{uR}$. To be able to
predict the size of such contributions, we shall assume that the up
quark mass matrix $M_u$ is approximately given by the texture zero
structure \cite{ricardo},
\begin{equation}
M_u = \left(\begin{array}{ccc}
0 & a & 0 \\
a & b & c \\
0 & c & d
\end{array}
\right)\ . \label{Mutexture}
\end{equation}
In this case
\begin{equation}
U_{uL} \sim U_{uR} \sim \left( \begin{array}{ccc} 1 & \sqrt{m_u/m_c}
& \sqrt{\epsilon m_u m_c^2/m_t^3} \\
-\sqrt{m_u/m_c} & 1 & \sqrt{\epsilon m_c/m_t}\\
\sqrt{\epsilon m_u/m_t} & -\sqrt{\epsilon m_c/m_t} & 1
\end{array}
\right)\ , \label{Uu}
\end{equation}
where
\begin{equation}
\epsilon \equiv \frac{d-m_t}{m_c}\ .
\end{equation}
Such a choice is of course in agreement with our previous assumption
of the matrices $U_{uL}$ and $U_{uR}$ being close to the identity
matrix.

Under the above conditions, the coupling matrix $G_3^u$ defined in
Eq.\,(\ref{G3u}) takes the simple form
\begin{equation}
G_3^u = g_t \left( \begin{array}{ccc} \epsilon m_u/m_t & -\epsilon
\sqrt{m_u m_c}/m_t & \sqrt{\epsilon m_u/m_t}\\
-\epsilon \sqrt{m_u m_c}/m_t & \epsilon m_c/m_t & -\sqrt{\epsilon
m_c/m_t}\\
\sqrt{\epsilon m_u/m_t} & -\sqrt{\epsilon m_c/m_t} & 1
\end{array}
\right)\ . \label{G3ua}
\end{equation}

In the context of our model, where the hierarchy $v \ll \sigma_b \ll
\sigma_t$ is expected among the VEV's, Eqs.\,(\ref{matrixO}),
(\ref{Au}) and (\ref{Ad}), together with (\ref{G3ua}) and
(\ref{G2da}), yield
\begin{align}
&A_1^u \simeq \frac{\sqrt{2}}{v}\, V^{\dagger}\, e^{i \varphi_b}
\left(
\begin{array}{ccc} 0 & 0 & 0 \\
0 & m_c(1-\epsilon\, e^{-\varphi_t})&
-\sqrt{\epsilon m_c m_t}\,e^{-i\varphi_t}\\
0 &-\sqrt{\epsilon m_c m_t}\,e^{-i\varphi_t} & m_t(1-e^{-i
\varphi_t})
\end{array} \right)\ ,\label{A1u} \\
&(A_1^d)_{ij} \simeq \frac{\sqrt{2}}{v}\,e^{i
\varphi_t}\,\sum_{k=1}^3\, V_{ik} \left[ (D_d)_{kj} - m_b V_{3k}^*
V_{3j} e^{-i \varphi_b} \right]\ ,
\label{A1d}\\
&A_2^u \ll A_1^u\ ,\quad A_2^d \ll A_1^d\ ,\label{A2u}
\end{align}
after the corresponding identification $v_1 = v, v_2 = \sigma_b\, ,
v_3 =\sigma_t$ and $\alpha_2 = -\varphi_b\,, \alpha_3 = \varphi_t\,$.
In particular, this implies that the contributions to
flavour-changing processes coming from the charged Higgs $H_2^+$ will
be strongly suppressed, provided that the Higgs mass $m_{H_2^+}
\simeq m_{H_1^+}$. In what follows we assume that the latter
condition is satisfied. Moreover, we shall discuss two limiting
cases: $\epsilon \simeq m_u/m_c \simeq 0$, which corresponds to $b
\simeq m_c$ in Eq.\,(\ref{Mutexture}), and $\epsilon \simeq 1$, i.e.
$b \simeq m_u \simeq 0$.

\section{New physics and $\mathbf{\ \varepsilon_K,\ \Delta m_{B_d},\
\Delta m_{B_s},\ \Delta m_D}$}
\label{sec:4}

Within the SM, the CKM matrix is constrained by unitarity and
experimental data. These constraints are usually expressed in terms
of the Wolfenstein parameters $A,\ \rho$ and $\eta$
\cite{wolfenstein}, and presented as a unitarity triangle in the
complex plane $(\bar{\rho},\bar{\eta})$ (see Fig.\,\ref{fig1} below)
\cite{PDG,gustavo}. They can be summarized as follows
\cite{buras,nir}:

From semileptonic $K$ and $B$ decays we have\footnote{All our input
parameters are taken from \cite{PDG,buras}.}
\begin{align}
&|V_{us}| = \lambda = 0.2205\ {\pm}\ 0.0018,\quad |V_{cb}| = 0.040\ {\pm}\
0.002,\nonumber\\
&|V_{ub}| = (3.56\ {\pm}\ 0.56) {\times} 10^{-3},\quad A =
\frac{|V_{cb}|}{\lambda^2} = 0.826\ {\pm}\ 0.041,
\end{align}
which implies
\begin{equation} \label{Rb}
R_b \equiv \sqrt{\bar{\rho}^2+\bar{\eta}^2} = \frac{1}{\lambda}
\left| \frac{V_{ub}}{V_{cb}} \right| = 0.39\ {\pm}\ 0.07\ ,
\end{equation}
with $\bar{\rho} = \rho(1-\lambda^2/2)\ ,\, \bar{\eta} =
\eta(1-\lambda^2/2)\ $. The above results are extracted from tree
level decays with large branching ratios and therefore their
determination is essentially independent of physics beyond the SM.

Next, for the $CP$ violating parameter $\varepsilon_K$ (and assuming
$\varepsilon_K \gg \varepsilon^\prime$),
\begin{equation}
\varepsilon_K \simeq \frac{e^{i \pi /4}}{\sqrt{2}} \frac{\im
\left(M_{12}^K \right)}{\Delta m_K}\ , \quad M_{12}^K = \frac{\langle
K^0 | {\cal H}_{\text{eff}}(\Delta S = 2) | \bar{K}^0 \rangle}{2
m_K}\ , \label{epsilon}
\end{equation}
the calculation of the box diagrams describing the $K^0 - \bar{K}^0$
mixing in SM gives
\begin{align} \label{epsilonSM}
&\varepsilon_K^{\text{SM}} = e^{i \pi /4} C_\varepsilon \hat{B}_K
\im(\lambda_t)[\re(\lambda_c^*)(\eta_1 S_0(x_c) - \eta_3
S_0(x_c,x_t))-\re(\lambda_t^*)\eta_2 S_0(x_t)]\ ,\nonumber\\
& \nonumber\\
&C_\varepsilon = \frac{G_F^2 m_W^2 f_K^2 m_K}{6 \sqrt{2} \Delta m_K}
= 3.84 {\times} 10^4\ ,\quad \lambda_i = (V_{is}^* V_{id})\ ,\quad x_i =
\frac{m_i^2}{m_W^2}\ .
\end{align}
Comparing this result with the experimental value
\mbox{$|\varepsilon_K| = (2.280\ {\pm}\ 0.013){\times} 10^{-3}$}, one obtains a
constraint in the form of the hyperbola
\begin{equation} \label{epsilonK}
\bar{\eta} \left[(1 -\bar{\rho})A^2\eta_2
S_0(x_t)+P_c(\epsilon)\right] A^2 \hat{B}_K = 0.226\ .
\end{equation}
In the above formulas, $P_c(\epsilon) = 0.31\ {\pm}\ 0.05$ summarizes the
charm-charm and charm-top contributions in the SM, $\hat{B}_K = 0.80\
{\pm}\ 0.15$ is a nonperturbative parameter, the correction factors
$\eta_1 = 1.38\ {\pm}\ 0.20,\ \eta_2 = 0.57\ {\pm}\ 0.01,\ \eta_3 = 0.47\ {\pm}\
0.04$ describe the short-distance QCD effects, $f_K = 160$~MeV is the
kaon decay constant, $m_K = 497.672\ {\pm}\ 0.031$ MeV is the kaon mass
and $\Delta m_K = (3.489\ {\pm}\ 0.008) {\times} 10^{-12}$~MeV is the mass
difference in the $K$ system. The gauge independent functions $S_0$
which govern the FCNC processes are approximately given by
\begin{align}
&S_0(x_t) = 2.46 \left(\frac{m_t}{170\,\gev}\right)^{1.52}\
,\quad S_0(x_c) = x_c\ ,\nonumber\\
&S_0(x_c,x_t) = x_c \left[\ln \frac{x_t}{x_c} -\frac{3 x_t}{4
(1-x_t)} - \frac{3 x_t^2 \ln x_t}{4 (1-x_t)^2} \right]\ ,
\end{align}
where $m_c = 1.30\ {\pm}\ 0.05$~GeV and $m_t = 165\ {\pm}\ 5$~GeV correspond
to the running quark masses defined as $m_q \equiv
m_q(m_q^{\text{pole}})$.

Substituting the numerical values for the parameters in
Eq.\,(\ref{epsilonK}) the $\varepsilon_K$ constraint reads
\begin{equation}\label{eK}
  \bar{\eta} [(1 -\bar{\rho})(0.91\ ^{+\ 0.16}_{-\ 0.14} + (0.31\ {\pm}\
  0.05)]  = 0.41\ ^{+\ 0.15}_{-\ 0.10}\ .
\end{equation}

Next, the amplitude for the $\Delta B =2$ transition in the
$B^0_{d,s} - \bar{B}^0_{d,s}$ systems is given in the SM by
\begin{align}
M_{12}^{\text{SM}}(B_q) &= \frac{\langle B^0_q | {\cal
H}_{\text{eff}}(\Delta B = 2) | \bar{B}^0_q \rangle}{2 m_{B_q}} =
\kappa_q (V_{tb} V_{tq}^*)^2\ ,\quad q = d,s\ ,\label{M12BqSM} \\
\kappa_q &= \frac{G_F^2}{12 \pi^2} m_W^2 \eta_B m_{B_q} f^2_{B_q}
\hat{B}_{B_q}  S_0(x_t)\nonumber \ ,
\end{align}
so that the mass differences are
\begin{equation} \label{DMBq}
\Delta m_{B_q} = 2 |M_{12}^{\text{SM}}(B_q)| = 2 |\kappa_q (V_{tb}
V_{tq}^*)^2|\ .
\end{equation}
Here $\eta_B = 0.55\ {\pm}\ 0.01$ is a QCD correction coefficient,
$m_{B_d} = 5.28$~GeV and $m_{B_s} = 5.37$~GeV are the $B$-meson
masses and the factor $f_{B_q} \hat{B}_{B_q}^{1/2}$ measures the
hadronic uncertainties. Recent lattice QCD estimates give $f_{B_d}
\hat{B}_{B_d}^{1/2} = 200\ {\pm}\ 40$~MeV and $\xi_s \equiv (f_{B_s}
\hat{B}_{B_s}^{1/2})/(f_{B_d} \hat{B}_{B_d}^{1/2}) = 1.14\ {\pm}\ 0.08$.

Combining the experimental value $\Delta m_{B_d} = 0.471\ {\pm}\
0.016\,{\text{ps}}^{-1}$ with Eq.\,(\ref{DMBq}) we can determine the
parameter
\begin{equation} \label{Rt}
R_t \equiv \sqrt{(1-\bar{\rho})^2+\bar{\eta}^2} = \frac{1}{\lambda}
\left| \frac{V_{td}}{V_{cb}} \right| = \left[\frac{|V_{td}|}{8.8 {\times}
10^{-3}}\right] \left[\frac{0.040}{|V_{cb}|} \right] = 0.98\ ^{+\
0.37}_{-\ 0.22}\ .
\end{equation}
On the other hand, the measurement $\Delta m_{B_s} > 12.4
\,{\text{ps}}^{-1}$ allows us to determine $R_t$ in a different way,
namely,
\begin{equation} \label{Rs}
R_t = \frac{1}{\lambda} \left| \frac{V_{td}}{V_{ts}} \right| =
\frac{\xi_s}{\lambda} \sqrt{\frac{m_{B_s}}{m_{B_d}}}
\sqrt{\frac{\Delta m_{B_d}}{\Delta m_{B_s}}} < 1.03\ ^{+\ 0.15}_{-\
0.14}\ .
\end{equation}
Fig.\,1 summarizes the constraints given by Eqs.\,(\ref{Rb}),
(\ref{eK}), (\ref{Rt}) and (\ref{Rs}) in the plane ($\bar{\rho},
\bar{\eta})$ within the SM. The dot-filled area corresponds to the
presently allowed region if no new physics beyond the SM is invoked.

\begin{figure}[htb]
$$\includegraphics[width=11cm]{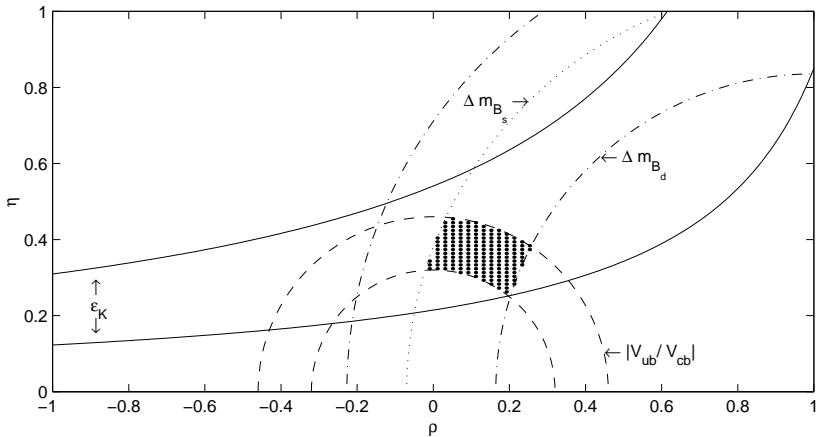}$$
\caption{Constraints on the plane ($\rho, \eta$) coming from the
measurements of $|V_{ub}/V_{cb}|$ (dashed), $\varepsilon_K$ (solid),
$\Delta m_{B_d}$ (dot-dashed) and $\Delta m_{B_s}$ (dotted) within
the Standard Model. The dot-filled area corresponds to the presently
allowed region.} \label{fig1}
\end{figure}

The model of interest to us and presented in Section \,\ref{sec:2}
contains new physical fields when compared to the SM. As the masses
of such fields are expected to be much larger than $m_W$, their
contributions to charged current tree level decays should be
negligible. They can however significantly contribute to quantities
such as $\varepsilon_K$ and $\Delta m_{B_{d,s}}$, thus playing an
important role in the determination of the unitarity triangle. To
establish their impact, first we shall compute the new contributions
to $\varepsilon_K, \Delta m_{B_d}$ and $\Delta m_{B_s}$ coming from
the FCNC processes induced by the heavy neutral Higgs field. Then we
shall compare these contributions with the ones induced by the new
heavy charged Higgs fields. As we shall see, if the mass scale for
the heavy charged Higgs ($m_{H^+}$) is of the same order than the
scale for the heavy neutral Higgs ($m_{H^0}$), new physics
contributions are always dominated by tree-level FCNC effects.
Finally, new contributions to the $\Delta m_D$ mass difference are
also expected and they are discussed at the end of this section.

\subsection{FCNC contributions}

Since the couplings of down quarks to the neutral scalar fields $R$
and $I$ are flavour-violating (cf. Eqs.\,(\ref{LYd}) and (\ref{Nd})),
they will induce a tree-level FCNC contribution to $K^0-\bar{K}^0$
mixing. In the framework of our model such couplings are determined
by Eqs.\,(\ref{Nd}), (\ref{G2da}) and given by
\begin{equation}
\Gamma^d_{ij} = -\frac{\sqrt{v^2+\sigma_b^2}}{v}\, e^{i \varphi_b}\,
(G_2^d)_{ij}\ \simeq -\frac{\sqrt{2}\, m_b}{v}\, e^{i
\varphi_b}\,V^\ast_{3i} V_{3j}\ , \label{G2dij}
\end{equation}
where we have used the fact that $v \ll \sigma_b ,\, m_b \simeq g_b
\,\sigma_b/ \sqrt{2}$. To estimate the hadronic matrix elements it is
customary to use the so-called vacuum insertion approximation
\cite{gustavo}. In this approximation, the new physics contribution
to the matrix element $M_{12}$ of the transition $K^0-\bar{K}^0$ will
be given by
\begin{equation}
M_{12}^{\text{new}}(K) = \frac{<K^0| {\cal
H}_{\text{eff}}^{\text{new}}|\bar{K}^0>}{2m_K} = - e^{2i\varphi_b} \,
\frac{f_K^2 m_K \hat{B}_K m_b^2}{4 v^2 m_{H^0}^2}\,(V_{td}^*
V_{ts})^2 \left[ \frac{1}{6}+\left(\frac{m_K}{m_d+m_s}\right)^2
\right]\ , \label{M12K}
\end{equation}
where ${\cal H}_{\text{eff}}^{\text{new}}$ is the effective $\Delta S
= 2$ Hamiltonian induced by the neutral Higgs exchange. It is now
straightforward to compute the FCNC contribution to the
$CP$-violating parameter $\varepsilon_K$ defined in
Eq.\,(\ref{epsilon}). We obtain
\begin{equation}
\varepsilon_K^{H^0} = -C^{(0)}_\varepsilon |V_{us}|^2 |V_{cb}|^4
\left\{ \left[(1-\rho)^2-\eta^2\right] \sin 2\varphi_b + 2 \eta
(1-\rho) \cos 2\varphi_b\right\} \frac{m_b^2}{m_{H^0}^2} \ ,
\label{epsilon0}
\end{equation}
where
\begin{equation}
C^{(0)}_{\varepsilon} = \frac{m_K f_K^2 \hat{B}_K}{4 \sqrt{2}\,\Delta
m_K v^2} \left[\frac{1}{6}+\left(\frac{m_K}{m_d+m_s}\right)^2\right]
\simeq 6.8 {\times} 10^{12} \left[\frac{\gev}{v}\right]^2\ ,
\end{equation}
with $m_s(m_c) = 130$~MeV, $m_d(m_c) = 8$~MeV. Substituting the
central values $|V_{us}| = 0.2205,\, |V_{cb}| = 0.040,\, m_b =
4.25$~GeV in Eq.\,(\ref{epsilon0}) and assuming $\varphi_b \simeq
\pi/2$, we find
\begin{equation}
\varepsilon_K^{H^0} \simeq 30.6\, \eta\, (1-\rho)
\left[\frac{\gev}{v}\right]^2 \left[\frac{\tev}{m_{H^0}}\right]^2\ .
\label{epsilonK0}
\end{equation}

A lower bound on the scale of the heavy neutral Higgs can be then
obtained by requiring the new physics contribution (\ref{epsilonK0})
to be smaller than the SM contribution, i.e. $|\varepsilon_K^{H^0}| <
|\varepsilon_K^{\text{SM}}|$. Since for the central values of the
parameters, the contribution to $\varepsilon_K$ in the SM (cf.
Eq.\,(\ref{epsilonSM})) is approximately given by
\begin{equation}
|\varepsilon_K^{\text{SM}}| \simeq 5.2 {\times} 10^{-3} \eta\, (1.34 -
\rho)\ , \label{epsilonSMc}
\end{equation}
we find for $0 \lesssim \rho \lesssim 0.3$,
\begin{equation}
m_{H^0} \gtrsim 65\, \tev\left[\frac{\gev}{v}\right]\ .
\label{boundK}
\end{equation}
In particular, for $v = \sqrt{2}\,m_c \simeq 1.84$~GeV we obtain
\begin{equation}
m_{H^0} \gtrsim 35\, \tev\ . \label{boundK0}
\end{equation}

New FCNC contributions induced by the heavy neutral Higgs field $H^0$
in the mass differences $\Delta m_{B_q}\, (q=d,s)$ are easily
obtained from the previous results on $\varepsilon_K$. For the matrix
elements $M_{12}^{\text{new}}(B_q)$ we have
\begin{align}
&M_{12}^{\text{new}}(B_q) =  - e^{2i\varphi_b} \,\kappa_q^{(0)}
\,(V_{tq}^* V_{tb})^2 \frac{m_b^2}{m_{H^0}^2}\ ,\label{M12Bq}\\
&\kappa_q^{(0)} = \frac{f_{B_q}^2 \hat{B}_{B_q} m_{B_q}}{4 v^2}
\left[ \frac{1}{6}+\left(\frac{m_{B_q}}{m_q+m_b}\right)^2 \right]\
,\nonumber
\end{align}
with
\begin{equation}
\kappa_d^{(0)} \simeq 0.09\,\gev \left[\frac{\gev}{v}\right]^2\
,\quad \kappa_s^{(0)} \simeq 0.12\,\gev
\left[\frac{\gev}{v}\right]^2\ .
\end{equation}
This implies
\begin{align}
\Delta m_{B_d}^{H^0}&=2 |M_{12}^{\text{new}}(B_d)|= 2 \kappa_d^{(0)}
|V_{us}|^2 |V_{cb}|^2  \left[(1-\rho)^2+\eta^2 \right]
\frac{m_b^2}{m_{H^0}^2}\nonumber\\
&\simeq 3.84 {\times} 10^2\, {\text{ps}}^{-1} \left[(1-\rho)^2+\eta^2
\right] \left[\frac{\gev}{v}\right]^2
\left[\frac{\tev}{m_{H^0}}\right]^2\ ,\\
\Delta m_{B_s}^{H^0}&= 2 |M_{12}^{\text{new}}(B_s)|= 2 \kappa_s^{(0)}
|V_{cb}|^2 \frac{m_b^2}{m_{H^0}^2} \simeq 1.02 {\times} 10^4\,
{\text{ps}}^{-1} \left[\frac{\gev}{v}\right]^2
\left[\frac{\tev}{m_{H^0}}\right]^2\ .
\end{align}

These FCNC contributions to $\Delta m_{B_q}$ are to be compared with
the SM contributions given by Eq.\,(\ref{DMBq}) and which can be
approximately written as
\begin{equation}
\Delta m_{B_d}^{\text{SM}} \simeq 0.48\,{\text{ps}}^{-1}
\left[(1-\rho)^2+\eta^2 \right]\ ,\quad \Delta m_{B_s}^{\text{SM}}
\simeq 13.04\,{\text{ps}}^{-1}\ .\label{DmBqSM}
\end{equation}
We have then
\begin{equation}
w_q \equiv \frac{\Delta m_{B_q}^{H^0}}{\Delta m_{B_q}^{\text{SM}}} =
\frac{\kappa_q^{(0)}}{\kappa_q} \frac{m_b^2}{m_{H^0}^2} \simeq 0.19
\left[ \frac{\gev}{v}\right]^2 \left[
\frac{65\,\tev}{m_{H^0}}\right]^2 \lesssim 0.19\ , \label{boundBq}
\end{equation}
if the lower bound given in Eq.\,(\ref{boundK}) for $m_{H^0}$ is
satisfied. We see that the contributions to $\Delta m_{B_q}$ coming
from the neutral Higgs are much smaller than the SM ones. In Fig.\,2
we illustrate our results for $m_{H^0} = 35$ TeV and $v=\sqrt{2}\,
m_c$. We notice that while $\varepsilon_K$ is quite sensitive to new
physics, the constraints coming from $B^0-\bar{B}^0$ mixing
practically do not change when compared to the SM results.

\begin{figure}[htb]
$$\includegraphics[width=11cm]{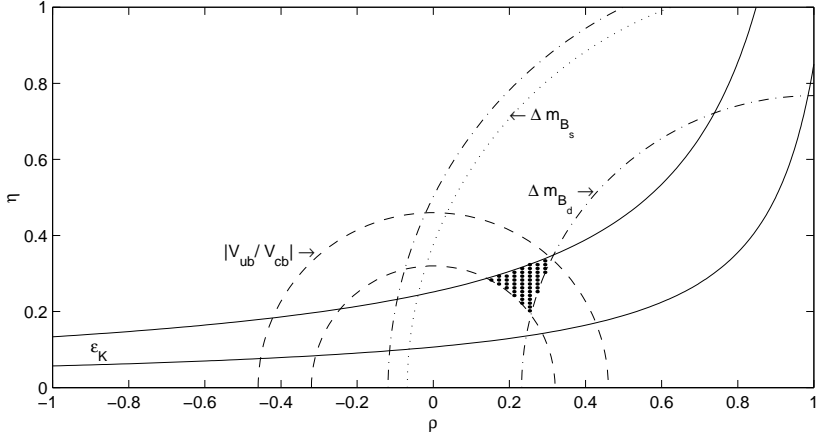}$$
\caption{Constraints on the plane ($\rho,\eta$) after including the
new FCNC contributions induced by the heavy neutral Higgs $H^0$. The
curves are given for $m_{H^0} =35$~TeV, $v = \sqrt{2}\, m_c$ and we
assume $m_{H_1^+} \simeq m_{H_2^+} \gg m_{H^0}$\,. The dot-filled
area corresponds to the region allowed by the present experimental
bounds.} \label{fig2}
\end{figure}

\subsection{Charged-current contributions}

The main contributions to flavour-changing processes induced by the
charged Higgs field $H_1^+$ are described by box diagrams where
$H_1^+$ and the $W$ gauge boson are circulating inside the
box\footnote{As discussed at the end of Section \ref{sec:3}, the
contributions coming from the charged Higgs field $H_2^+$ can be
neglected if $m_{H_2^+} \simeq m_{H_1^+}$ (cf. Eq.\,(\ref{A2u})).}.
Their computation is also straightforward.

For the new physics contribution to the amplitude $M_{12}$ in the
$K^0-\bar{K}^0$ system we find
\begin{align}
M_{12}^{\text{new}}(K) &= \frac{\sqrt{2}\,G_F m_W^2 m_K f_K^2
\hat{B}_K}{24 \pi^2 v^2} \left[ \lambda_t^* \Gamma_{tt} f_t +
\lambda_c^* \Gamma_{cc} f_c + (\lambda_{ct}^*
\Gamma_{ct}+\lambda_{tc}^* \Gamma_{tc})
 f_{ct}\right] \frac{m_t^2}{m_{H^+}^2} \ ,\nonumber\\
f_t &= -\frac{x_t}{2}\left[1 + \ln x_t - \frac{3}{x_t-1} + \frac{3
\ln
x_t}{(x_t-1)^2}\right]\ , \quad f_c = -2x_c\,(1+\ln x_c)\ ,\nonumber\\
f_{ct} &= \sqrt{x_c x_t} \left[\frac{2x_c \ln
x_c}{x_t}+\frac{3}{2}\,\frac{\ln x_t}{x_t-1} + \frac{1}{2} \ln
\frac{x_{H^+}}{x_t} \right]\ ,\nonumber\\
\Gamma_{ij} &= g_{id}\, g_{js}^*\ ,\quad \lambda_{ij} =
V_{is}^*V_{jd}\ ,\quad i=c,t\ ,\label{M12KH+}
\end{align}
with $\lambda_i$ and $x_i$ defined in Eq.\,(\ref{epsilonSM}).
Moreover, according to Eq.\,(\ref{A1u}),
\begin{align}
&g_{t\alpha} = -V_{c\alpha}^*\, e^{-i \varphi_t} \sqrt{\frac{\epsilon
m_c}{m_t}}
+ V_{t\alpha}^* (1-e^{-i \varphi_t})\ ,\label{gij}\\
&g_{c\alpha} = V_{c\alpha}^* (1-\epsilon\, e^{-i \varphi_t})
\frac{m_c}{m_t}  - V_{t\alpha}^*\,e^{-i \varphi_t}
\sqrt{\frac{\epsilon m_c}{m_t}} \ ,\quad \alpha = d,s,b\ .\nonumber
\end{align}

Therefore, the charged Higgs contribution to $\varepsilon_K$ reads
\begin{equation}
\varepsilon_K^{H^+} = C_\varepsilon^{(+)} \left[A_t f_t + A_c f_c +
A_{ct} f_{ct}\right] \frac{m_t^2}{m_{H^+}^2} \ ,\label{epsilon+}
\end{equation}
where
\begin{align}
&A_t = \im (\lambda_t^* \Gamma_{tt})\ ,\quad A_c = \im (\lambda_c^*
\Gamma_{cc})\ ,\quad A_{ct} = \im (\lambda_{tc}^* \Gamma_{tc} +
\lambda_{ct}^* \Gamma_{ct})\ ,\nonumber\\
&C_\varepsilon^{(+)} =\frac{G_F m_W^2 m_K f_K^2 \hat{B}_K}{24 \pi^2
v^2 \Delta m_K} \simeq 0.93 {\times} 10^9
\left[\frac{\gev}{v}\right]^2\ .
\end{align}
When the texture parameter $\epsilon=0\ $, the coefficients $A_i$ in
Eq.\,(\ref{epsilon+}) are approximately given by
\begin{align}
&A_t \simeq 4 (1 - \cos \varphi_t) |V_{cb}|^2 (1- \rho) J\ ,\quad
A_c \simeq 0\ ,\nonumber\\
&A_{ct} \simeq 2 \frac{m_c}{m_t} (1 - \cos \varphi_t) J\ ,\quad J =
|V_{us}|^2 |V_{cb}|^2 \eta\ ,
\end{align}
while for $\epsilon = 1$ they are
\begin{equation}
A_t \simeq A_c \simeq J \frac{m_c}{m_t}\ ,\quad A_{ct} \simeq 2 J (1
- \rho)\frac{m_c}{m_t}\ .
\end{equation}

We can now give a numerical estimate of the above contributions.
Using the fact that $\cos \varphi_t \simeq 1 - m_b^2/(2 m_t^2)$ we
obtain
\begin{align}
&\left. \varepsilon_K^{H^+}\right|_{\epsilon = 0} \simeq -0.016\,
\eta\,(1-\rho)
\left[\frac{\gev}{v}\right]^2 \left[\frac{\tev}{m_{H^+}}\right]^2\ ,\\
&\left. \varepsilon_K^{H^+}\right|_{\epsilon = 1} \simeq -2.55\,
\eta\,(23.6 + \rho) \left[\frac{\gev}{v}\right]^2
\left[\frac{\tev}{m_{H^+}}\right]^2\ .
\end{align}
If we require that $|\varepsilon_K^{H^+}| <
|\varepsilon_K^{\text{SM}}|$\,, where $\varepsilon_K^{\text{SM}}$ is
given in Eq.\,(\ref{epsilonSMc}), then we can find the following
lower bounds on the charged Higgs mass for $0 \lesssim \rho \lesssim
0.3$,
\begin{align}
&m_{H^+} \gtrsim 1.5\,\tev \left[\frac{\gev}{v}\right]\quad
\text{for}\quad \epsilon=0\ ,\label{boundK+e=0}\\
&m_{H^+} \gtrsim 101\,\tev \left[\frac{\gev}{v}\right]\quad
\text{for}\quad \epsilon=1\ .\label{boundK+e=1}
\end{align}
This in particular implies for $v \simeq \sqrt{2}\,m_c\ $,
\begin{equation}
m_{H^+} \gtrsim 800\,\gev \quad \text{if}\quad \epsilon=0\ ,\quad
m_{H^+} \gtrsim 55\,\tev \quad \text{if}\quad \epsilon=1\ .
\end{equation}

Let us now consider the charged Higgs contributions to
$B_{d,s}^0-\bar{B}_{d,s}^0$ mixing. Their computation is analogous to
the one in the $K^0-\bar{K}^0$ system. We obtain
\begin{align}
&M_{12}^{\text{new}}(B_q) = \kappa_q^{(+)} \left[A_{tt}^{(q)} f_t +
A_{cc}^{(q)} f_c + (A_{ct}^{(q)}+A_{tc}^{(q)}) f_{ct}\right]
\frac{m_t^2}{m_{H^+}^2} \ ,\label{M12BqH+}\\
&\kappa_q^{(+)} = \frac{\sqrt{2}\,G_F m_W^2 m_{B_q} f_{B_q}^2
\hat{B}_{B_q}}{24 \pi^2 v^2}\ ,\quad A_{ij}^{(q)} = V_{ib} V_{jq}^*
g_{iq} g_{jb}^*\ ,\quad q=d,s\ ,\nonumber
\end{align}
with
\begin{equation}
\kappa_d^{(+)} \simeq 0.95 {\times} 10^{-4}\, \gev
\left[\frac{\gev}{v}\right]^2\ ,\quad \kappa_s^{(+)} \simeq 1.26
{\times} 10^{-4}\, \gev \left[\frac{\gev}{v}\right]^2\ ,
\end{equation}
$f_i$ and $g_{ij}$ defined in Eqs.\,(\ref{M12KH+}) and (\ref{gij}),
respectively.

For $\epsilon=0$ the amplitude (\ref{M12BqH+}) is dominated by the
top quark contributions described by the coefficient $A_{tt}^{(q)}$.
We have
\begin{equation}
A_{tt}^{(q)} \simeq 2 (V_{tb} V_{tq}^*)^2\,(1-\cos \varphi_t) \simeq
(V_{tq}^*)^2 \frac{m_b^2}{m_t^2}\ .
\end{equation}
The top-charm terms proportional to $A_{tc}^{(q)}$ will however give
an important contribution to the amplitude in the case of $\epsilon
=1$ . In the latter case we find for the relevant coefficients:
\begin{align}
&A_{tt}^{(q)} \simeq V_{tq}^* V_{cq}^*
V_{tb}^2\,(1-e^{-i\varphi_t})\sqrt{\frac{m_c}{m_t}} \simeq i V_{tq}^*
V_{cq}^* \frac{m_b}{m_t} \sqrt{\frac{m_c}{m_t}}\ ,\nonumber\\
&A_{tc}^{(q)} \simeq (V_{cq}^*)^2 V_{tb}^2 \frac{m_c}{m_t} \simeq
(V_{cq}^*)^2 \frac{m_c}{m_t}\ .
\end{align}

Therefore, the new contributions to the mass differences $\Delta
m_{B_q}$ will be given by
\begin{align}
\left. \Delta m_{B_d}^{H^+}\right|_{\epsilon=0} &= 2 \kappa_d^{(+)}
|V_{us}|^2 |V_{cb}|^2 \left[(1-\rho)^2+\eta^2\right] |f_t|
\frac{m_b^2}{m_{H^+}^2}\nonumber\\
&\simeq 1.65\,{\text{ps}}^{-1}\left[(1-\rho)^2+\eta^2\right]
\left[\frac{\gev}{v}\right]^2
\left[\frac{\tev}{m_{H^+}}\right]^2\ ,\\
\left. \Delta m_{B_s}^{H^+}\right|_{\epsilon=0} &= 2 \kappa_s^{(+)}
|V_{cb}|^2 |f_t| \frac{m_b^2}{m_{H^+}^2} \simeq
22.4\,{\text{ps}}^{-1}\left[\frac{\gev}{v}\right]^2
\left[\frac{\tev}{m_{H^+}}\right]^2\ .
\end{align}
Comparing these values with the SM result given in
Eqs.\,(\ref{DmBqSM}) we arrive at
\begin{equation}
\frac{\Delta m_{B_q}^{H^+}}{\Delta m_{B_q}^{\text{SM}}} \simeq 1.55
\left[ \frac{\gev}{v}\right]^2 \left[
\frac{1.5\,\tev}{m_{H^+}}\right]^2 \ .\label{boundBq+}
\end{equation}
We see that for $m_{H^+}$ close to the lower bound (\ref{boundK+e=0})
the charged Higgs contributions to $B_{d,s}^0-\bar{B}_{d,s}^0$ mixing
are of the same order of magnitude than the SM ones. A slightly
higher value of $m_{H^+}$ is required if we impose $\Delta
m_{B_q}^{H^+}<\Delta m_{B_q}^{\text{SM}}$. From Eq.\,(\ref{boundBq+})
it follows then
\begin{equation}
m_{H^+} \gtrsim 1\,\tev \quad \text{for}\quad \epsilon=0\ ,
\label{boundBd+e=0}
\end{equation}
with $v \simeq \sqrt{2}\,m_c\ $.

In the case of $\epsilon=1$ we have
\begin{align}
\left. \Delta m_{B_d}^{H^+}\right|_{\epsilon=1} &= 2 \kappa_d^{(+)}
|V_{us}|^2 \left|-i V_{cb} (1-\rho + i
\eta)\sqrt{\frac{m_c}{m_t}}\,f_t + \frac{m_c}{m_b}\, f_{ct} \right|
\frac{m_b m_t}{m_{H^+}^2}\nonumber\\
&\simeq 140\,{\text{ps}}^{-1}\left|1.8-\eta+0.7\ln
\left(\frac{m_{H^+}}{\tev}\right)+i (1-\rho)\right|
\left[\frac{\gev}{v}\right]^2
\left[\frac{\tev}{m_{H^+}}\right]^2\ ,\\
\left. \Delta m_{B_s}^{H^+}\right|_{\epsilon=1} &= 2 \kappa_s^{(+)}
 \left|-i V_{cb} \sqrt{\frac{m_c}{m_t}}\,f_t + \frac{m_c}{m_b}\, f_{ct} \right|
\frac{m_b m_t}{m_{H^+}^2}\nonumber\\
&\simeq 3.8 {\times} 10^3\,{\text{ps}}^{-1}\left|1.8+0.7\ln
\left(\frac{m_{H^+}}{\tev}\right)+i \right|
\left[\frac{\gev}{v}\right]^2 \left[\frac{\tev}{m_{H^+}}\right]^2\ .
\end{align}

Comparing these contributions with the SM result (\ref{DmBqSM}) we
find:
\begin{equation}
\frac{\Delta m_{B_q}^{H^+}}{\Delta m_{B_q}^{\text{SM}}} \lesssim 0.15
\ , \label{boundBq+e=1}
\end{equation}
for $m_{H^+} \gtrsim 55\,\tev$ and $v \simeq \sqrt{2}\,m_c\ $ as
given by the bound (\ref{boundK+e=1}).

In Fig.\,3 we illustrate the constraints on the ($\rho,\eta$)-plane
for $m_{H^+} = 2$~TeV, $v = \sqrt{2}\, m_c$ and the parameter
$\epsilon = 0$. We assume $m_{H^0} \gg m_{H^+}$ and thus, only the
contribution coming from the flavour-changing charged current is
taken into account. The dot-filled area corresponds to the allowed
region. A similar plot is given in Fig.\,4 for the case $\epsilon =
1$ and $m_{H^+} = 80$~TeV.

\begin{figure}[htb]
$$\includegraphics[width=11cm]{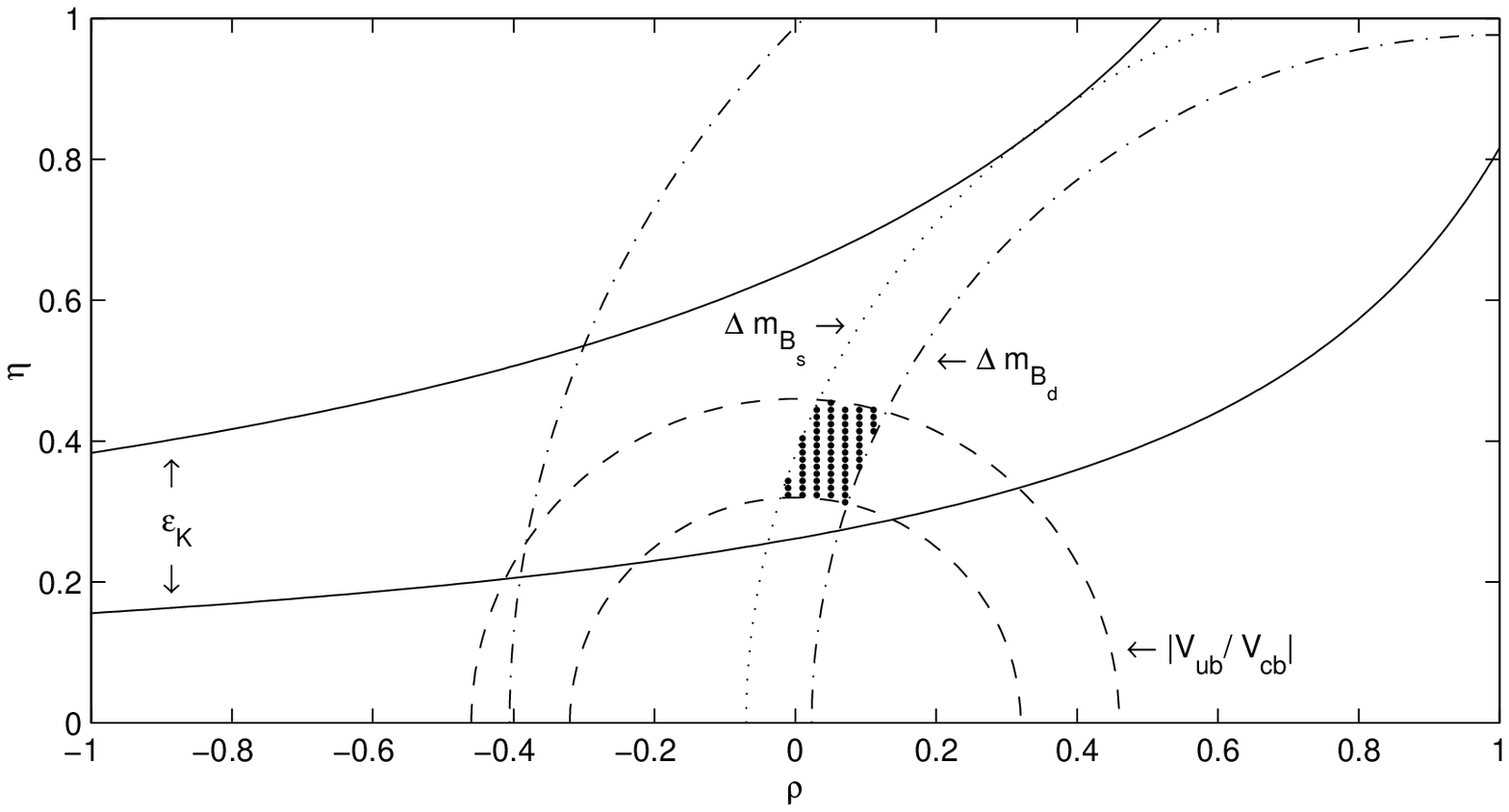}$$
\caption{Constraints on the plane ($\rho,\eta$) after including the
new flavour-changing charged contributions induced by the heavy
charged Higgs $H_1^+$. The curves are given for $m_{H_1^+} =2$~TeV,
$v = \sqrt{2}\, m_c$ and the parameter $\epsilon = 0$. We assume
$m_{H_1^+} \simeq m_{H_2^+} \ll m_{H^0}$\,.} \label{fig3}

$$\includegraphics[width=11cm]{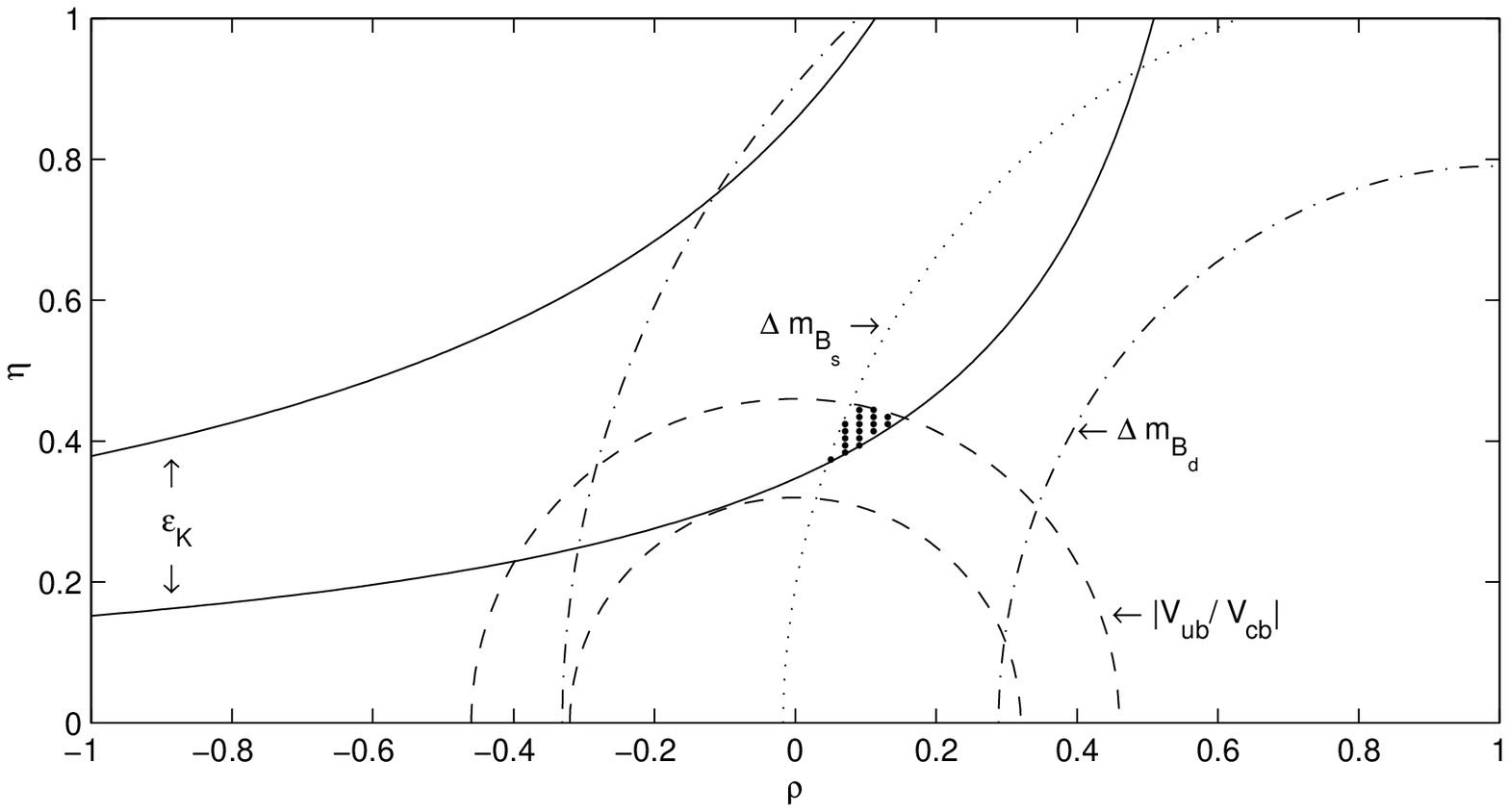}$$
\caption{As in Fig.\,3, but taking the parameter $\epsilon = 1$ and
$m_{H_1^+} = 80$~TeV.} \label{fig4}
\end{figure}

\subsection{FCNC and $\mathbf{\Delta m_D}$}

In the up quark sector, $D^0-\bar{D}^0$ mixing is perhaps one of the
most interesting processes given that this process is highly
suppressed in the SM: $\Delta m_D^{\text{SM}} < 10^{-15}$~GeV. To
estimate the size of FCNC contributions to the mass difference
$\Delta m_D$ we use the expression (\ref{Nu}) and the approximate
form (\ref{Uu}) for the matrices $U_{uL}, U_{uR}$. The relevant
couplings are then given by
\begin{equation}
\Gamma_{ij}^u = -\frac{\sigma_b \sigma_t}{v^2}e^{-i\varphi_t}
(G_3^u)_{ij} \simeq -\frac{2 m_b m_t}{v^2}e^{-i\varphi_t}
(G_3^u)_{ij}\ , \label{G3uij}
\end{equation}
where $G_3^u$ is given in Eq.\,(\ref{G3ua}) and we have used $m_b
\simeq g_b \sigma_b/\sqrt{2}\ $, $m_t \simeq g_t \sigma_t/\sqrt{2}$,
$g_{b,t} \simeq 1$, to approximate the right hand side of
Eq.\,(\ref{G3uij}).

Keeping the dominant term we obtain
\begin{align}
M_{12}^{\text{new}}(D) & = - \epsilon^2 e^{-2i \varphi_t}
\kappa_u^{(0)}
\frac{m_b^2}{m_{H^0}^2}\ ,\\
\kappa_u^{(0)} & = \frac{4 m_u m_c f_D^2 \hat{B}_D m_D}{v^4} \left[
\frac{1}{6}+\left(\frac{m_D}{m_u+m_c}\right)^2 \right]\ ,\nonumber\\
& \simeq 5.4 \times 10^{-3}\, \gev \left[\frac{\gev}{v}\right]^4\
,\nonumber
\end{align}
where $f_D \sqrt{\hat{B}_D} \simeq 225$ MeV \cite{UKQCD}, $m_D =
1.86$ GeV and we take $m_u \simeq 5\,\mev$. Therefore,
\begin{equation}
\Delta m_D = 2 |M_{12}(D)| = 2 \epsilon^2 \kappa_u^{(0)}
\frac{m_b^2}{m_{H^0}^2} \simeq 4.6\, \epsilon^2 \times 10^{-11}\,\gev
\left[\frac{\gev}{v}\right]^4
\left[\frac{65\,\tev}{m_{H^0}}\right]^2\ .
\end{equation}
Using the bound on $m_{H^0}$ coming from $K$ physics (cf.
Eq.\,(\ref{boundK0})), one obtains the upper limit
\begin{equation}
\Delta m_D \lesssim 1.35\, \epsilon^2 {\times} 10^{-11}\, \gev\ .
\end{equation}

We notice that except when $\epsilon \ll 1$, our model predicts a
value for $\Delta m_D$ much larger than in the SM. This is a clear
signature of the model.

Comparing the above value with the experimental limit
\begin{equation}
\left(\Delta m_D \right)_{\text{exp}} < 5 {\times} 10^{-14}\,\gev\ ,
\end{equation}
we find an upper limit on the parameter $\epsilon$,
\begin{equation}
\epsilon \lesssim 0.06\ , \label{epsilonmD}
\end{equation}
which in turn translates into constraints on the texture form
(\ref{Mutexture}) assumed here for the up quark mass matrix $M_u$.

At this point it is worth recalling that flavour-changing
contributions induced by new physics crucially depend on the specific
patterns of the fermion mass matrices. In our analysis we have
assumed a simple but quite generic \emph{Ansatz} for the up-quark
mass matrix, expressed in terms of quark mass ratios and a free
parameter $\epsilon$. This parameter is therefore expected to be
constrained in order to avoid dangerous large contributions to
low-energy observable effects. It is clear that more specific mass
matrix textures (e.g. triangular-type textures in the up-quark
sector) could lead to a natural suppression of these contributions
and the above constraints thus be avoided.

\section{Electric dipole moment of the neutron}

The electric dipole moment (EDM) of a fermion $\Psi$ is defined as
$$
iF_{D}(k^2)\overline{\Psi}(p_1)\gamma _5\sigma ^{\mu \nu}\Psi
(p_2)F_{\mu \nu}\ ,
$$
where $F_{\mu \nu}$ is the electromagnetic tensor and $k_\mu =
p_{2\mu}-p_{1\mu}$. In the present model, we can expect a large
contribution to the EDM of the neutron induced by the spontaneous
$CP$ violation if the phases $\varphi_{t,b} \neq 0,\pi $. In the
latter case, the exchange of the heavy neutral and the heavy charged
Higgs fields are expected to contribute to the EDM of the neutron.
For the down quark contribution with exchange of a heavy neutral
Higgs the dominant term is given by ($f=d,u$)
\begin{equation}
\left| \frac{F_D(0)}{e}\right|_f \simeq \frac{|Q_f|}{16\pi^2}\,
\im(\Gamma_{31}^f\Gamma_{13}^f) \frac{\sqrt{x_f} \,(x_f^2-1-2x_f \ln
x_f)}{(x_f-1)^3\,m_{H^0}}\ , \label{edmH0d}
\end{equation}
where $x_{d,u} \equiv m_{b,t}^2/m_{H^0}^2\ $, $Q_u=2/3\ $, $Q_d=-1/3\
$, $\Gamma _{ij}^d$ and $\Gamma _{ij}^u$ are defined in
Eqs.\,(\ref{G2dij}) and (\ref{G3uij}), respectively. Using the fact
that $m_b \simeq g_b \sigma_b/\sqrt{2}\ $, $m_t \simeq g_t
\sigma_t/\sqrt{2}\ $ and assuming as before $g_{b,t} \simeq 1$, we
have
\begin{align}
&\Gamma_{31}^d \simeq -\frac{\sqrt{2}\, m_b}{v}\, e^{i
\varphi_b}\,V_{td}\ ,\quad  \Gamma_{13}^d \simeq -\frac{\sqrt{2}\,
m_b}{v}\, e^{i\varphi_b}\,V_{td}^*\ ,\\
&\Gamma_{31}^u = \Gamma_{13}^u \simeq -\frac{2 m_b \sqrt{\epsilon m_u
m_t}}{v^2} e^{-i\varphi_t}\ .
\end{align}
Since $x_f \ll 1$ and $\sin 2\varphi_{b,t} \simeq 2 m_b/m_t$,
Eqs.\,(\ref{edmH0d}) are approximately given by
\begin{align}
\left| \frac{F_D(0)}{e}\right|_d &\simeq \frac{1}{12 \pi^2}
\left|V_{td}\right|^2 \frac{m_b^4}{v^2 m_t m_{H^0}^2} \simeq 8 {\times}
10^{-30}\, \text{cm} \left[\frac{\gev}{v}\right]^2
\left[\frac{65\,\tev}{m_{H^0}}\right]^2\ ,
\\
\left| \frac{F_D(0)}{e}\right|_u &\simeq  \frac{\epsilon}{3\pi^2}
 \frac{m_u m_b^3 m_t}{v^4 m_{H^0}^2} \simeq \epsilon {\times}
10^{-23}\, \text{cm} \left[\frac{\gev}{v}\right]^4
\left[\frac{65\,\tev}{m_{H^0}}\right]^2\ ,
\end{align}
after substituting the values $m_u \simeq 5\,\mev, m_b \simeq
4.25\,\gev, m_t \simeq 165\,\gev$ and $|V_{td}| \simeq 0.01$.

Thus, using the lower bound on $m_{H^0}$ given by
Eq.\,(\ref{boundK0}), we obtain the following upper bounds on the
electric dipole moments of the quarks:
\begin{equation}
\left| \frac{F_D(0)}{e}\right|_d \lesssim 8 {\times} 10^{-30}\, \text{cm}\ ,
\quad \quad \left| \frac{F_D(0)}{e}\right|_u \lesssim 3 \,\epsilon {\times}
10^{-24}\, \text{cm}\ . \label{edmdu}
\end{equation}

Taking into account that $m_u/m_c \simeq 10^{-3} \lesssim \epsilon
\leq 1$ and assuming
\begin{equation}
\left| \frac{F_D(0)}{e}\right|_n \simeq \left| \frac{F_D(0)}{e}
\right|_d+\left|\frac{F_D(0)}{e}\right|_u\ ,
\end{equation}
we conclude from Eqs.\,(\ref{edmdu}) that the EDM will be always
dominated by the up-quark contribution. Moreover,
\begin{equation}
10^{-27}\, \text{cm} \lesssim \left| \frac{F_D(0)}{e}\right|_n
\lesssim 10^{-24}\, \text{cm}
\end{equation}
in the allowed range of the parameter $\epsilon$. Of course, the
above conclusions hold for our specific choice of the up-quark mass
matrix texture (\ref{Mutexture}). If such is the case, then the
predicted EDM is expected to be very close to the present
experimental limit \cite{PDG}
\begin{equation}
\left| \frac{F_D(0)}{e}\right|_{\text{exp}} < 10^{-26}\, \text{cm}\ .
\label{edmexp}
\end{equation}

The comparison of Eqs.\,(\ref{edmdu}) with the experimental bound
(\ref{edmexp}) allows us to further constrain the parameter
$\epsilon$. We find
\begin{equation}
10^{-3} \lesssim \epsilon \lesssim 3 \times 10^{-3}\ ,
\label{epsilonEDM}
\end{equation}
which is more restrictive than the upper bound previously found from
$D^0-\bar{D}^0$ mixing (see Eq.\,(\ref{epsilonmD})). Notice however
that the constraint (\ref{epsilonEDM}) was obtained assuming the
lower bound on the Higgs mass given by Eq.\,(\ref{boundK0}) and
implied by $\varepsilon_K$. We could of course relax this constraint
by pushing the heavy neutral Higgs mass to a higher scale, but then
the ``raison d'\^{e}tre" of our model would be lost and its predictions
would be close to the SM ones. From a phenomenological point of view
we find more plausible to fix the Higgs scale from the constraints
coming from $K$ and $B$ physics and, in particular, from the
$CP$-violating parameter $\varepsilon_K$. Other constraints, such as
the EDM of the neutron, will then give us a hint on what kind of
quark mass matrix textures are favoured in the theory.

Let us now consider the charged Higgs contributions. In this case the
dominant contribution is coming from the diagrams with a top quark
circulating inside the loop. The photon line can be attached to the
Higgs line or to the quark line. Therefore we have
\begin{align}
\left| \frac{F_D(0)}{e}\right|_{H^+} &\simeq \frac{1}{16\pi^2} \im
\left[(A_1^d)_{31}(A_1^u)_{13}\right]  \left\{
Q_{H^-}\frac{\sqrt{x_u}\,(x_u^2-1-2x_u \ln x_u)}{(x_u-1)^3}\right.
\nonumber \\
&+ \left. Q_u\frac{\sqrt{x_u}\,(x_u^2 - 4x_u + 3 + 2 \ln
x_u)}{(x_u-1)^3} \right\}\,\frac{1}{m_{H^+}}\ ,
\end{align}
with $x_u$ now defined as $x_u \equiv m_t^2/m_{H^+}^2\ $,
$Q_{H^-}=-1$; $A_1^u$ and $A_1^d$ given by Eqs.\,(\ref{A1u}) and
(\ref{A1d}), respectively. Since
\begin{align}
(A_1^u)_{13} &\simeq \frac{\sqrt{2}}{v} e^{i\varphi_b} V_{td}^* m_t
(1-e^{i\varphi_t}) \simeq -\frac{i\sqrt{2}}{v}\,
e^{i\varphi_b}V_{td}^* m_b\ , \\
(A_1^d)_{31} &\simeq - \frac{\sqrt{2}}{v}\,
e^{i(\varphi_t-\varphi_b)}\,V_{td} m_b\ ,
\end{align}
we obtain
\begin{equation}
\left| \frac{F_D(0)}{e}\right|_{H^+} \simeq \frac{3}{8 \pi^2}
\frac{m_b^2 m_t}{v^2 m_{H^+}^2}\left|V_{td}\right|^2 \simeq 2 \times
10^{-22}\, \text{cm} \left[\frac{\gev}{v}\right]^2
\left[\frac{\tev}{m_{H^+}}\right]^2\ .  \label{edmH+}
\end{equation}

Imposing the experimental constraint given in Eq.\,(\ref{edmexp}), we
get a stronger constraint on $m_{H^+}$ than the lower bound
(\ref{boundK+e=0}), namely,
\begin{equation}
m_{H^+} \gtrsim 140\,\tev \left[ \frac{\gev}{v}\right]\ .
\end{equation}
In particular, for $v \simeq \sqrt{2}\, m_c$ we obtain
\begin{equation}
m_{H^+} \gtrsim 75\,\tev\ . \label{bound edm}
\end{equation}
We also remark that in leading order this bound is independent of the
texture parameter $\epsilon\ $.

\section{FCNC in top decays: the example $t \rightarrow q\, \gamma$}

The FCNC decays $t\rightarrow q\, \gamma $ and $t \rightarrow q\,Z$
are strongly suppressed in the SM at the level of $10^{-12}$.
Observation of any of these events would be an indication of physics
beyond the SM. It is of particular interest to study the order of
magnitude that our model predicts for such processes. The amplitude
of $t\rightarrow q\,\gamma $ can be parametrized as
\begin{equation}
M_{\gamma } \equiv \bar{q}(p_1)\left[ i(A + B \gamma_5)\sigma^{\mu
\nu }\frac{q_\nu}{m_t}\right] t(p_2)A_\mu\ ,
\end{equation}
where $A_\mu$ is the photon field and $q_\mu \equiv
p_{2\mu}-p_{1\mu}$. The decay width of this process is given by
\begin{equation}
\Gamma (t\rightarrow q\,\gamma)=\frac{m_t}{8\pi}\left( \left|
A\right|^2+\left|B\right|^2\right)\ ,
\end{equation}
and is dominated by the process $t \rightarrow b\,W$,
\begin{equation}
\Gamma (t\rightarrow b\,W) = \frac{G_F}{8\sqrt{2}\,\pi}\left|
V_{tb}\right|^{2}m_t^3\,\left(1-\frac{m_W^2}{m_t^2}\right) \left(1 +
\frac{m_W^2}{m_t^2}-2 \frac{m_W^4}{m_t^4}\right)\ .
\end{equation}

Once the coefficients $A$ and $B$ are known, it is straightforward to
compute the branching ratio corresponding to the process $t
\rightarrow q\, \gamma\ $. Recently, an experimental limit on this
process has been reported \cite{CDF}
\begin{equation}
B(t\rightarrow q\gamma ) < 0.032\ ,
\end{equation}
which can be translated into a limit on the parameters $A$ and $B$,
\begin{equation}
\left| A\right| ^{2}+\left| B\right| ^{2} < 6.5 \times\, 10^{-3}\ .
\end{equation}
This limit should improve with the future LHC, which is expected to
decrease the above bound by two orders of magnitude \cite{beneke}
\begin{equation}
B(t \rightarrow q\,\gamma ) < 10^{-4}\ ,
\end{equation}
i.e.
\begin{equation}
|A|^2+|B|^2 < 2 {\times} 10^{-5}\ .
\end{equation}

In our model one expects that a such process will be induced by one
loop diagrams as is the case for the EDM. Moreover, both
contributions, with a heavy neutral Higgs exchange and with a charged
Higgs exchange, should be taken into account. One expects then using
the bounds given by Eqs.\,(\ref{boundK0}) and (\ref{bound edm}),
\begin{equation}
A \approx B \simeq \frac{m_{b}^{2}}{m_{H^{0}}^{2}}\lesssim 10^{-8}\
\end{equation}
for the neutral Higgs exchange, while
\begin{equation}
A \approx B \simeq \frac{m_{t}^{2}}{ m_{H^{+}}^{2}}\lesssim 5 \times
10^{-6}\
\end{equation}
for the charged Higgs exchange.

Since the above bounds are further multiplied by additional
suppression factors coming from the CKM matrix, we can conclude that
the prediction of the present model for the process $t \rightarrow
q\,\gamma $ is out of the reach of the next future colliders.

\section{Implications on \textsl{CP} asymmetries}

As discussed in Section\,\ref{sec:4}, if $m_{H^+} \simeq m_{H^0}$
then the flavour-changing charged Higgs contributions to $\Delta
m_{B_{d,s}}$ are negligible. In order to study the implications of
the model on $CP$ asymmetries, we shall assume that new physics
appears only through tree-level FCNC effects.

An easy way to parametrize the effects of new physics on
$B_q^0-\bar{B}_q^0$ mixing is by introducing the parameters $r_q^2$
and the phases $2\theta_q$\, through the relation

\begin{equation}
M_{12}(B_q) = M_{12}^{\text{SM}}+M_{12}^{\text{new}} \equiv r_q^2
e^{2i \theta_q} M_{12}^{\text{SM}}(B_q)\ . \label{M12rq}
\end{equation}

On the other hand, from Eqs.\,(\ref{M12BqSM}) and (\ref{M12Bq}) we
have
\begin{equation}
M_{12}(B_q) =  ( 1-w_q e^{2i\varphi_b}) M_{12}^{\text{SM}}(B_q)\ ,
\label{M12wq}
\end{equation}
with $w_q$ defined in Eq.\,(\ref{boundBq}). Comparing
Eqs.\,(\ref{M12rq}) and (\ref{M12wq}) we find the relations
\begin{equation}
r_q^2 = \sqrt{1-2w_q \cos 2\varphi_b + w_q^2}\ ,\quad \tan 2 \theta_q
 = \frac{ -w_q \sin 2 \varphi_b }{1-w_q \cos 2 \varphi_b}
\label{thetaq}\ .
\end{equation}

Within the SM, the $CP$ asymmetry $a_{\psi K_s}$ in $B_d^0\,
(\bar{B}_d^0) \rightarrow \psi K_S\ $ decays is related to the angle
$\beta$ of the unitarity triangle,
\begin{equation}
a_{\psi K_s}= \sin 2 \beta\ ,\quad \beta = \arg \left[-\frac{V_{cd}
V_{cb}^*}{V_{td} V_{tb}^*}\right]\ .
\end{equation}
While global analyses of the CKM unitarity triangle yield the values
\begin{equation}
(\sin 2\beta)_{\text{SM}} = \left\{
\begin{array}{ll}
0.75 \pm 0.06 & ~ \cite{stocchi}\ , \\
0.73 \pm 0.20 & ~ \cite{ali}\ ,\\
0.63 \pm 0.12 & ~ \cite{schune}\ ,
\end{array}
\right.
\end{equation}
the recent experimental measurements of the above time dependent $CP$
asymmetry give
\begin{equation}
(\sin 2\beta)_{\psi K_S} =\left\{
\begin{array}{ll}
0.12 \pm 0.37 \pm 0.09 & \text{(BABAR)} ~\cite{BABAR}\ , \\
0.45 \pm 0.44 \pm 0.08 & \text{(BELLE)} ~~\cite{BELLE}\ , \\
0.79 \pm 0.42 & \text{(CDF)} ~~~~~\cite{CDF1}\ .
\end{array}
\right.
\end{equation}
The above experimental values imply the average
\begin{equation}
(\sin 2\beta)_{\psi K_S} = 0.42 \pm 0.24\ . \label{sinbetaexp}
\end{equation}
Although the SM estimates are consistent with the present
experimental results, the small values of $\sin 2\beta$ found by
BABAR and BELLE collaborations might indicate the presence of new
physics contributions.

If the new physics modifies the phase of the mixing amplitude, then
the asymmetry will also get a contribution from the $\theta_d$ phase,
\begin{equation}
a_{\psi K_s}= \sin 2 (\beta + \theta_d)\ .
\end{equation}

Moreover, if we assume that the $\theta$ term in Lagrangian
(\ref{eq:6}) is the only source of $CP$ violation in our model, then
$\varphi_b \simeq -\pi/2 + m_b/m_t\ $. In this case
Eqs.\,(\ref{thetaq}) imply
\begin{equation}
r_d \simeq \sqrt{1+w_d}\ ,\quad \tan 2 \theta_d \simeq \frac{2
w_d}{1+w_d} \frac{m_b}{m_t}\ .
\end{equation}
We see that the new phase $\theta_d$ is suppressed by the ratio
$m_b/m_t\ $. Using the upper bound $w_d \lesssim 0.19$ given by
Eq.\,(\ref{boundBq}) we find
\begin{equation}
r_d \lesssim 1.1\ , \quad \quad r_s \simeq r_d\ , \quad \quad \tan 2
\theta_d \lesssim 0.008\ . \label{boundtd}
\end{equation}
Although these predictions are consistent with the global average
(\ref{sinbetaexp}), we notice that the deviations from the SM
predictions are very small in this case and, consequently, it is not
possible to achieve consistency \cite{nir} with the small values
reported by the BABAR collaboration \cite{BABAR}.

In order to illustrate the dependence of our result on the strong
$CP$ phase $\theta$, let us assume that $CP$ violation in the CKM
matrix is independent of the value of $\theta$. In other words, let
us assume that besides the angle $\theta$, there exist other sources
of $CP$ violation and we consider $\theta$ as an arbitrary parameter.
Of course, when $\theta$ is different from $\pi/2$, the isospin
symmetry between the top and bottom quarks has to be explicitly
broken in the effective Lagrangian (\ref{eq:3}). From the
minimization of the potential (cf. Eqs.\,(\ref{eq:10})) and taking
$m_b \ll m_t$, one easily finds
\begin{equation}
\varphi_b \simeq - \theta + \frac{m_b}{m_t}\ , \quad \varphi_t \simeq
\frac{m_b}{m_t}\ .
\end{equation}
At first order in $m_b/m_t$ we obtain in this case:
\begin{equation}
r_d = \sqrt{1-2 w_d \cos 2\theta +w_d^2}\ ,\quad \tan 2\theta_d =
\frac{w_d \sin 2 \theta}{1- w_d \cos 2 \theta}\ .
\end{equation}

It is interesting to study how the angle $\theta_d$ varies as a
function of $\theta$. The extrema values for $\theta_d$ are obtained
when $\cos 2\theta = w_d$ and this implies
\begin{equation}
r_d = \sqrt{1-w_d^2}\ ,\quad \tan 2\theta_d = \pm
\frac{w_d}{\sqrt{1-w_d^2}}\ .
\end{equation}
With $w_d \lesssim 0.19$ as given by Eq.\,(\ref{boundBq}) we have
then
\begin{equation}
r_d \gtrsim 0.98\ ,\quad  |\tan 2\theta_d| \lesssim 0.19\ .
\end{equation}
Thus, if the strong $CP$ phase $\theta$ is assumed to be a free
parameter of the model, the constraint (\ref{boundtd}) on the new
physics contribution to the $CP$ asymmetries in $B$ decays is relaxed
and we can reach a rather sizeable phase $\theta_d \simeq 6^\circ$.
This in turn would allow to accommodate \cite{nir} the present
experimental measurements, including the small values obtained by
BABAR and BELLE collaborations.

\section{Conclusion}

In this paper we have studied the phenomenological constraints on a
model where $CP$ violation is dynamically induced by a strong $CP$
phase $\theta$ \cite{delepine}. The most promising tests for the
model are given by the new experimental prospects to measure $\Delta
M_{D}$ or to improve the experimental limit on the electric dipole
moment of the neutron.

Contrary to naive expectations, the fact that the new force
responsible for the electroweak symmetry breaking and for the
spontaneous $CP$ violation is only sensitive to the third generation
of quarks does not imply that the most stringent constraints come
from processes involving the heavy flavours ($t$ and $b$). We have
shown that the stringent constraints on the scale of new physics come
from $K$ physics and from the electric dipole moment of the neutron.
This means that even if FCNC processes are naturally suppressed in
the model by the CKM matrix elements, this suppression is not strong
enough to allow for a mass scale of the heavy Higgs to be of the
order of few TeV.

\bigskip
\noindent \textbf{\large Acknowledgements}
\bigskip

We are grateful to L.T. Handoko, F. Kr{\"u}ger and A. Teixeira for
useful discussions and comments. One of us (R.G.F.) would like to
thank the {\em Funda\c{c}{\~a}o para a Ci{\^e}ncia e a Tecnologia} for
financial support under the grant SFRH/BPD/1549/2000. One of us
(D.D.) would like to thank the DESY theory group where part of this
work was completed.

\end{document}